\documentstyle[aps]{revtex}

\begin{document}

\title{Time symmetric action-at-a-distance electrodynamics \\ and the structure of SpaceTime}

\author{C\u alin Galeriu}

\address{Physics Department, Clark University, Worcester, MA, 01610, USA}

\maketitle

\begin{abstract}
This paper begins with a critical analysis of the concept of 'material point particle'. We argue that this concept is incompatible with the force laws of action-at-a-distance electrodynamics, and we suggest that the trajectory of a particle (the world-line) should be looked upon as a string in static equilibrium. By complementing our model with a time symmetric interaction law, we are led to a straightforward derivation of the principles underlying the electromagnetic interaction between two uniformly moving charged particles. An extension of our theory to the case of particles in arbitrary motion shows that we have to modify Maxwell's equations of the {\it microscopic} electromagnetic field, in order to accommodate a field-strength tensor which is no longer antisymmetric. It is further argued that the radiation reaction force gives another example of a 4-force which is not orthogonal to the 4-velocity, thus leading to a variable rest mass. The variation of the rest mass is then discussed from the point of view of the equivalence principle of general relativity, and we end up with a theoretical derivation of the homogenous Maxwell's equations.
\end{abstract}
   
\section{Introduction}

It is well known that the general solution of Maxwell's equations for the electromagnetic field is a sum of two contributions, one due to the
retarded potential and another due to the advanced potential \cite{davies}. Since the advanced solution does not seem to be observed in nature, several approaches have been developed to deal with this problem. Ritz \cite{ritz} assumes that the retarded solution of the electromagnetic field is of a more fundamental nature than Maxwell's equations from whom it initially originated. Dirac \cite{dirac} instead stresses the importance of ingoing (advanced) and outgoing (retarded) radiation fields as boundary conditions for Maxwell's differential equations. The main result of Dirac's work is a derivation of the radiative reaction force acting on an accelerated electron, with no reference to the internal structure of the electron. The theory however raises further questions, due to the introduction of acausal effects, and due to the {\it ad hoc} elimination of the divergence resulting from the interaction of the electron with its own generated field. Tetrode \cite{tetrode} has proposed to interpret the electromagnetic radiation as a consequence of an interaction between a source and an observer. The elimination of the electromagnetic field as a stand alone entity was not a new idea. Ritz \cite{ritz} has argued that the field is a mathematical artifact, since one can observe only the Lorentz force acting on charged particles. Relativistic action-at-a-distance electrodynamics considers the null cone of special relativity as the surface of interparticle action \cite{davies}. By eliminating the electromagnetic field one also eliminates the problem of the self-interaction divergence, and this is one of the main advantages of action-at-a-distance electrodynamics \cite{wheeler1}. Wheeler and Feynman \cite{wheeler2}, using time symmetric action-at-a-distance and field theory as equivalent and complementary tools, have  shown that, if we suppose a completely absorbing Universe, then all the advanced effects are canceled out except those which are comprised in the force of radiative reaction. They further  describe how, in spite of the time symmetric interaction used, the irreversibility of the emission process appears as a phenomenon of statistical mechanics connected with the asymmetry of the initial conditions with respect to time. The elaborations in this present paper, although not emerging directly from the work of the previously cited authors, are a further development of time symmetric action-at-a-distance electrodynamics.

\section{The world-line of a particle is a string in static equilibrium}

Consider the following paradoxical situation: {\it a force along the x-axis can modify the y-component of the velocity of a particle}. An example of such a situation is the motion of a relativistic charged particle (with a non-zero initial $y$-component of the velocity) in a uniform electric field parallel to the $x$-axis \cite{landau}. The origin of this situation lies in the fact that the magnitude of the 4-momentum is constant. The projection of the 4-momentum onto a 3D real space introduces into the expressions of the $x,y$ and $z$ components of the relativistic momentum a dependence on the $v_x,v_y$ and $v_z$ components of the velocity. Thus, while $v_x$ is affected more directly by the applied force, the $v_y$ component needs also to change (when $v_x$ changes), such that the $y$-component of the relativistic momentum remains constant. Since there is no {\it a priori} reason for the magnitude of the 4-momentum to be constant, one can at least ask whether there is a physical interpretation for this remarkable fact. The answer to the question is: yes, there is a perfect analogy between a string under constant tension and a world-line in 4D Minkowski space, between the tension in the string (constant in magnitude and always tangent to the string) and the 4-momentum (constant in magnitude and always tangent to the world-line). The analogy goes even further, since we can show that the equations of motion for a material point particle are equivalent with the static equilibrium condition for a 4D string. Indeed, consider two events, A and B, along the world-line of a particle, separated by an infinitesimal interval $ds = i \ c \ d\tau$, as shown in Figure~\ref{4Dstring}. Then the equation of motion of the particle
\begin{equation}
{\bbox{\cal P}_B - \bbox{\cal P}_A \over d\tau} = \bbox{\cal F}
\end{equation}
can be rewritten as the static equilibrium condition
\begin{equation}
\bbox{\cal T}_B + (-\bbox{\cal T}_A) + \bbox{f} \ ds = 0
\end{equation}
for a string under a tension $\bbox{\cal T} = i c \bbox{\cal P} / s_o = \bbox{\cal P} / \tau_o$ and acted upon by a linear force density $\bbox{f} = -\bbox{\cal F} / s_o$ (notice the change of sign, related to the change in direction of the force, as pictured in Figure~\ref{4Dstring} - we thus have to consistently swap attraction and repulsion when switching between the two models). For dimensional reasons we have introduced the constant $s_o = i c \tau_o$, heuristically called 'the length of a particle', measured along its world-line.\

Although at this moment the advantage of writing (2) instead of (1) is not evident, we will recall that the static description is closer to the mathematical structure of SpaceTime: "The objective world simply is; it does not happen. Only to the gaze of my consciousness, crawling upward along the life line [world-line] of my body, does a section of this world come to life as a fleeting image in space which continuously changes in time."\cite{weyl}

\section{Material point particles are incompatible with action-at-a-distance electrodynamics}

Two fundamental postulates are required to describe the electromagnetic interaction between two uniformly moving charged particles \cite{galeriu}. The first postulate is that in a reference frame where all the source charges producing a Coulombian electric field $\bbox{E}$ are at rest, the force on a charge $q$ is given by $\bbox{F} = q \bbox{E}$ independent of the velocity of the charge in that frame \cite{jackson}. The second postulate is the orthogonality of the 4-force to the 4-velocity, required in order to have the constant magnitude of the 4-momentum preserved. From this two postulates one can calculate the 4-force (in Gaussian units) with which a source charge $Q$, at rest at the origin, is acting on a test charge $q$, at a distance $\bbox{R}$ and moving with velocity $\bbox{v}$:
\begin{equation}
\bbox{\cal F}_{\rm real} = \bbox{\hat R} {Q q \over R^2} \gamma(v)
\end{equation}
\begin{equation}
\bbox{\cal F}_{\rm imag} = \bbox{\hat i} {Q q \over R^2} \gamma(v) {v_{\rm rad} \over c}
\end{equation}
where the radial component of the velocity is $v_{\rm rad} = \bbox{\hat R}\cdot\bbox{v}$ .

Although generally accepted, the above two postulates raise some questions. We usually think of a particle as being a point in Minkowski space: "We fix our attention on the substantial point which is at the world-point $x,y,z,t$, and imagine that we are able to recognize this substantial point at any other time." \cite{minkowski} 

But, from the local point of view, a point in Minkowski space is just a fixed point, it does not have a velocity! The velocity comes from the tangent to the world-line of the particle. Consequently we have to consider more that one point (one instance of the material point particle) on the world-line in order to observe the velocity. We are thus lead to the conclusion that the 4-force from (3)-(4), the relevant Minkowski vector of the problem, is incompatible with the material point particle model, since we have in the 4-force (3)-(4) an explicit dependence on the velocity of the test particle.
Another argument in favor of our conclusion is that one cannot define, in SpaceTime, a relativistically invariant distance between two interacting material point particles. This is because, due to the speed of light at which the interaction propagates, the SpaceTime interval between the two points is null. There is no privileged reference frame such that the two events be at the same place or happen at the same time (which would have allowed us to define a relativistically invariant scalar - the distance between the particles in the same-time frame, or the delay between the events in the same-place frame). The radius $R$ entering (3)-(4) is the direct result of the selection of the reference frame in which the source charge is at rest, but this, once again, requires us to consider {\it the velocity} of this particle. 

\section{An imaginative argument in favor of time symmetric electrodynamics}

A strange thing happens when we introduce the finite velocity of propagation of the electromagnetic interaction. While in the old instantaneous action model there is a strong electrostatic component directed along the line connecting the two interacting particles, this is no longer the case when we introduce the retardation time. It is no
longer intuitively clear why two attracting particles (in a system having no angular momentum) will end up in the same SpaceTime point. However, we might imagine the actual electromagnetic 4-force as being the sum of two 4-forces (the retarded and the advanced one, as shown in Figure~\ref{forces}) which are directed along the lines connecting the interacting particles. In the stationary case the imaginary components of these 4-forces just cancel, and the real components produce the Coulombian force. As we will see in the next section, this picture is correct, up to a change of sign for the imaginary 4-force components.

\section{An all-geometrical derivation for the electromagnetic interaction between two uniformly moving particles}

We will introduce now a physical model which is consistent with all the previous considerations. Instead of a material point particle with mass $m_o$ and charge $q$, we will consider a 4D string (the world-line) with linear mass density $\mu_o = m_o / s_o$ and linear charge density $\rho = q /s_o$. If particles are regarded as points, then each particle (one point) interacts with all other particles (points) which lie in the past (or future) part of its light cone. On the other hand, a particle of length $s_o$ (an infinitesimal length element along the world-line) interacts with all of the segments of the world-lines of other particles which are included between the light cones of the starting point of the segment $s_o$ and of the ending point of the same segment. Accordingly, the electrostatic force law $F = Qq / R^2$ between a point source charge $Q$ and a point test charge $q$ will be replaced by a 4-force ${\cal F} = (\rho_Q S_o) (\rho_q s_o) / R^2 = (Qq / R^2) (S_o / s_o)$, acting on a 'particle' of length $s_o$, and corresponding to an electrostatic linear 4-force density law $-\bbox{f} = \bbox{\cal F} / s_o$. Of course, since the linear force density is meant to act on an infinitesimal length element, the segment $s_o$ has to be very small (probably less than $10^{-15}$ cm), but still not zero, since it appears in the denominator of the linear mass, charge and 4-force densities. It is obvious that, if the two particles are at rest, or move with the same velocity, then $S_o = s_o$, and we recover the Coulombian electrostatic interaction. We will next show that, if there is a relative velocity between the two particles, then the $S_o / s_o$ length ratio is responsible for the dependence on the velocity of the 4-force given in (3)-(4). We will thus analyze the situation of a source particle at rest, and of a test particle at a position $\bbox{R}$ relative to the first one, moving with velocity $\bbox{v}$. 

\subsection{Length ratios for radial velocity}
Consider first the simpler case of only radial motion. We will use the same geometrical technique developed in \cite{galeriu}.
If the test particle (AB) has only a radial velocity $v$, it will 'see' a segment (FG) in the past of the source particle, and a segment (HI) in the future, as shown in Figure~\ref{radial}.
\begin{equation}
DC = CE = i \ CB = i \ AB \ \sin(-\alpha)
\end{equation}
\begin{equation}
AC = AB \ \cos(-\alpha)
\end{equation}
\begin{equation}
FG = AD = AC - DC = AB \ [ \cos(-\alpha) - i \sin(-\alpha) ] = AB \ \gamma(v) (1 - v/c)
\end{equation}
\begin{equation}
HI = AE = AC + CE = AB \ [ \cos(-\alpha) + i \sin(-\alpha) ] = AB \ \gamma(v) (1 + v/c)
\end{equation}
Thus the length ratio for the retarded part is $\gamma(v) (1 - v/c)$, and for the advanced part is $\gamma(v) (1 + v/c)$. 

\subsection{Length ratios for arbitrary velocity}
What happens if the test particle has a velocity, but not only in the radial direction? We will rotate our 3D reference frame in such a way that the radial direction gives the Ox direction, and that the velocity of the test particle lies in the (xOy) plane. The radial component of the velocity is therefore $v_y$. Consider a point A on the world-line of the second particle, as shown in Figure~\ref{notradial}. We have to find the points H and J at which the world-line of the source particle intersects the light cones of the point A. For this we draw, from the point A, a parallel to Oy, which intersects the (xOict) plane at B. From B, a parallel to Ox intersects the Oict axis at C, and the world-line of the source particle at F.
\begin{equation}
OC = EF = ict
\end{equation}
\begin{equation}
CB = v_x t
\end{equation}
\begin{equation}
BA = v_y t
\end{equation}
\begin{equation}
EO = FC = R
\end{equation}
\begin{equation}
OA = \sqrt{OC^2 + CB^2 + BA^2} = i t \sqrt{c^2 - v_x^2 - v_y^2}
\end{equation}
\begin{equation}
FA = \sqrt{BA^2 + (FC + CB)^2} = \sqrt{v_x^2 t^2 + (R + v_y t)^2}
\end{equation}
\begin{equation}
GE = EI = i \ EO = i R
\end{equation}
\begin{equation}
HF = FJ = i \ FA = i \sqrt{v_x^2 t^2 + (R + v_y t)^2}
\end{equation}
For the segment GH from the past:
\begin{equation}
GH = GF - HF = GE + EF - HF = i R + i c t - i \sqrt{v_x^2 t^2 + (R + v_y t)^2}
\end{equation}
For the segment IJ from the future:
\begin{equation}
IJ = EJ - EI = EF + FJ - EI = i c t + i \sqrt{v_x^2 t^2 + (R + v_y t)^2} - i R
\end{equation}
Assuming an infinitesimal length OA of the test particle, and using l'H\^opital rule, it follows that: 
\begin{equation}
lim_{A \to O} {GH \over OA} = lim_{t \to 0} {i R + i c t - i \sqrt{v_x^2 t^2 + (R + v_y t)^2}
\over i t \sqrt{c^2 - v_x^2 - v_y^2}} = {c - v_y \over \sqrt{c^2 - v_x^2 - v_y^2}} = \gamma(v) (1 - {v_{\rm rad} \over c})
\end{equation}
\begin{equation}
lim_{A \to O} {IJ \over OA} = lim_{t \to 0} {i c t + i \sqrt{v_x^2 t^2 + (R + v_y t)^2} - i R \over i t \sqrt{c^2 - v_x^2 - v_y^2}} = {c + v_y \over \sqrt{c^2 - v_x^2 - v_y^2}} = \gamma(v) (1 + {v_{rad} \over c})
\end{equation}
Thus the length ratio for the retarded part is $\gamma(v) (1 - v_{\rm rad}/c)$, and for the advanced part is $\gamma(v) (1 + v_{\rm rad}/c)$.

\subsection{The postulates of the new time symmetric action-at-a-distance electrodynamics}
Corresponding to the length ratios just derived, if we have a source particle of charge $Q$ (a world-line of linear charge density $Q/s_o$) at rest at the origin, a test particle of charge $q$ (a segment of length $s_o$ along a world-line of linear density $q/s_o$) will 'see' a charge $Q \gamma(v) (1 - v_{\rm rad}/c)$ in the past, and a charge $Q \gamma(v) (1 + v_{\rm rad}/c)$ in the future. We postulate that, in the reference frame in which the source particle is at rest, the 4-force acting on the test particle will have a real radial component proportional to the (-2) power of the distance $R$ on the real axis, and an imaginary component proportional to the (-2) power of the distance $i R$ on the imaginary axis. Consequently, attraction on the real axis is associated with repulsion on the imaginary axis, and {\it vice versa}. Both components of the 4-force will be proportional to the product of charges. We also postulate that we have interaction with both the future and the past. Because the two contributions will add, the proportionality constant in the Coulombian 4-force will be multiplied by $1/2$. This corresponds to adding one half of the retarded potential to one half of the advanced one, in a time symmetric electrodynamics \cite{davies}.

The 4-forces acting on the test particle, as shown in Figure~\ref{timesym}, will be:
\begin{equation}
\bbox{\cal F}_{\rm past,real} = \bbox{\hat R} {1 \over 2} {Q q \over R^2} \gamma(v) 
(1 - {v_{\rm rad} \over c})
\end{equation}
\begin{equation}
\bbox{\cal F}_{\rm past,imag} = - \bbox{\hat i} {1 \over 2} {Q q \over R^2} \gamma(v) 
(1 - {v_{\rm rad} \over c})
\end{equation}
\begin{equation}
\bbox{\cal F}_{\rm future,real} = \bbox{\hat R} {1 \over 2} {Q q \over R^2} \gamma(v) 
(1 + {v_{\rm rad} \over c})
\end{equation}
\begin{equation}
\bbox{\cal F}_{\rm future,imag} = \bbox{\hat i} {1 \over 2} {Q q \over R^2} \gamma(v) 
(1 + {v_{\rm rad} \over c})
\end{equation}
The total 4-force will have the real and imaginary parts:
\begin{equation}
\bbox{\cal F}_{\rm real} = \bbox{\cal F}_{\rm past,real} + \bbox{\cal F}_{\rm future,real} = 
\bbox{\hat R} {Q q \over R^2} \gamma(v)
\end{equation}
\begin{equation}
\bbox{\cal F}_{\rm imag} = \bbox{\cal F}_{\rm past,imag} + \bbox{\cal F}_{\rm future,imag} = 
\bbox{\hat i} {Q q \over R^2} \gamma(v) {v_{\rm rad} \over c}
\end{equation}
in agreement with (3) and (4).

We believe that the geometrical derivation of the Coulombian interaction between two charged particles that we have given brings us closer to Minkowski's vision: "[...] in my opinion physical laws might find their most perfect expression as reciprocal relations between these world-lines." \cite{minkowski}

Our theory of time symmetric action-at-a-distance electromagnetic interaction applies well to the case of uniformly moving charged particles. It is worth remembering that, although initially emerging from experiments with electrostatic charge distributions and stationary currents,  "Maxwell's equations are often used in a wider context, such as for accelerating charges, than the experimental evidence, on the basis of which they are normally developed." \cite{rosser}
Maxwell's equations for accelerated charges are thus checked only {\it a posteriori}. In a recently developed action-at-a-distance Gaussian electrodynamics \cite{domina} new tangential force components are also predicted for accelerated charges.

\section{The radiative reaction 4-force is parallel to the 4-velocity}
It is not hard to see that, if we apply our interaction model to accelerated charges, then the total 4-force, in general, will no longer be orthogonal to the 4-velocity of the test particle. It can be shown that, in general, our relativistic Lorentz 4-force is still given by ${\cal F}_{\alpha} = (q/c) T_{\alpha\beta} {\cal V}_{\beta}$, with the only difference that the field-strength tensor $T_{\alpha\beta}$ is no longer antisymmetric. We will try first to answer the fundamental question of whether a 4-force can be not orthogonal to the 4-velocity. 

This problem has appeared long ago, when the ponderomotive 4-force, in a system which dissipates energy by Joule heating, was considered \cite{pauli}. Abraham \cite{abraham} has shown that, since an inertial mass must be ascribed to every kind of energy ($E = m_o c^2$), the {\it rest mass} of the system has to decrease, corresponding to the Joule heat dissipated. The ponderomotive 4-force must thus have a component parallel to the 4-velocity, and the equations of motion are modified accordingly:
\begin{equation}
\bbox{\cal F} = {d \over d\tau} (m_o \bbox{\cal V}) = m_o {d \bbox{\cal V} \over d\tau} + \bbox{\cal V} {d m_o \over d\tau}
\end{equation}
It is worth mentioning that equation (27) can still be interpreted as the static equilibrium conditions for a 4D string, however this time under a variable tension (the tension is proportional to the rest mass).
From (27), the rate of energy dissipation is given by the 4-force component parallel to the 4-velocity:
\begin{equation}
(\bbox{\cal F} \cdot \bbox{\cal V}) = - c^2 {d m_o \over d\tau} = - \gamma(v) {d E \over dt}
\end{equation}

It is questionable why the same approach has not been applied to the case of radiation damping.
The radiative reaction force is introduced in order to satisfy an energy balance, for the nonrelativistic situation first \cite{jackson}. Thus the work done by the radiative reaction force has to equal the energy dissipated through electromagnetic radiation:
\begin{equation}
\int_{t_1}^{t_2} (\bbox{F}_{\rm rad} \cdot \bbox{v}) dt = - {2 \over 3} {q^2 \over c^3} \int_{t_1}^{t_2} (\bbox{\dot v} \cdot \bbox{\dot v}) dt \Rightarrow {2 \over 3} {q^2 \over c^3} \int_{t_1}^{t_2} (\bbox{\ddot v} \cdot \bbox{v}) dt
\end{equation}
The last part of (29) results from the Larmor power formula (34), if we integrate by parts, and assume that the motion is either periodic, or $(\bbox{\dot v} \cdot \bbox{v}) = 0$ at the moments $t_1$ and $t_2$. The radiation reaction force extracted this way is thus somehow averaged, and it does not reflect the instantaneous damping force.

In a first questionable step, from (29) the radiative reaction force is extracted as:
\begin{equation}
\bbox{F}_{\rm rad} = {2 \over 3} {q^2 \over c^3} \bbox{\ddot v}
\end{equation}
We have to warn that, since in (29) $\bbox{F}_{\rm rad}$ is in scalar product with $\bbox{v}$, the only meaningful information that can be extracted is about the component of the force which is parallel to the velocity, $(\bbox{F}_{\rm rad} \cdot \bbox{v}) \bbox{v} / v^2$!
Another problem related to the expression (30) is that it is not clear whether the force is indeed a {\it damping} force, pointing in the opposite direction than the velocity. This problem is evident if we consider the 'runaway' solution \cite{jackson}, in which the velocity, the acceleration and the acceleration's derivative are all parallel, pointing in the same direction, and increasing exponentially. 

In a second questionable step, the force from (30) is generalized \cite{davies,landau,pauli} to the relativistic case by introducing the derivative with respect to the proper time, and by adding an extra term, specifically needed to ensure the orthogonality between the 4-force and the 4-velocity. The relativistic 4-force obtained this way is:
\begin{equation}
\bbox{\cal F}_{\rm rad} = {2 \over 3} {q^2 \over c^3} [{d^3 \bbox{\cal X} \over d\tau^3} + {1 \over c^2} ({d^2 \bbox{\cal X} \over d\tau^2} \cdot {d^2 \bbox{\cal X} \over d\tau^2}) {d \bbox{\cal X} \over d\tau}]
\end{equation}
Since the only reason for being of the radiation reaction force is to account for the dissipation of energy, and this dissipation is accompanied by a decrease in the rest mass of the system, and furthermore only the component of the force parallel to the velocity enters the equation, we can safely consider the radiation reaction 4-force as being parallel to the 4-velocity. Therefore, there is no need for the last term in (31).

By extracting, from the first term in (31), the component parallel to the 4-velocity, and by requiring that this force point in the opposite direction than the 4-velocity, we obtain:
\begin{equation}
\bbox{\cal F}_{\rm rad} = {2 \over 3} {q^2 \over c^3} |{d^3 \bbox{\cal X} \over d\tau^3} \cdot {d \bbox{\cal X} \over d\tau}| {1 \over - c^2} {d \bbox{\cal X} \over d\tau}
\end{equation}
From (28) and (32) we can calculate the rate of energy dissipation:
\begin{equation}
{d E \over dt} = {- 1 \over \gamma(v)} (\bbox{\cal F}_{\rm rad} \cdot {d \bbox{\cal X} \over d\tau}) = {- 1 \over \gamma(v)} {2 \over 3} {q^2 \over c^3} |{d^3 \bbox{\cal X} \over d\tau^3} \cdot {d \bbox{\cal X} \over d\tau}| = {- 1 \over \gamma(v)} {2 \over 3} {q^2 \over c^3} |{d^2 \bbox{\cal X} \over d\tau^2} \cdot {d^2 \bbox{\cal X} \over d\tau^2}|
\end{equation}
In the nonrelativistic limit we recover the exact (not averaged!) Larmor power formula:
\begin{equation}
{d E \over dt} = - {2 \over 3} {q^2 \over c^3} a^2
\end{equation}
Therefore we claim that (32) is the correct expression of the instantaneous radiation reaction force.

\section{On a macroscopic scale, the 4-force component parallel to the 4-velocity cancels}
We now make the following conjecture: that the {\it averaged} electromagnetic 4-force (Lorentz force + radiative reaction) is still orthogonal to the 4-velocity. Our conjecture is based on the detailed analysis of a simple situation, shown in Figure~\ref{velchange}. A source particle moves with uniform radial velocity up to a point A, where its velocity is changed, and the uniform radial motion continues afterwards. During this time a test particle is held at rest by other external forces that exactly cancel the 4-force due to the source particle. It is evident that, before the point B or after the point D, the test particles 'sees' a uniformly moving source, with no apparent change of velocity between the retarded and advanced positions. Therefore, only on the segment BD can the electromagnetic 4-force have a component parallel to the 4-velocity of the test particle. We will calculate the total change of the rest mass (of the tension), on the segment BD, and we will show that there is no net change. For this, consider the interaction with the advanced source, as shown in Figure~\ref{advanced}. A particle at point P will be subject to a real 4-force and to an imaginary 4-force, given in the proper reference frame of the source particle. Thus the total component parallel to the 4-velocity of the test particle will be:
\begin{equation}
{\cal F}_{\Vert} = i {\cal F} \cos(-\alpha) - {\cal F} \sin(\alpha) = i {\cal F} {c \over \sqrt{c^2 - v^2} } - {\cal F} {i v \over \sqrt{c^2 - v^2} } = i {\cal F} \gamma(v) (1 - {v \over c})
\end{equation}
The magnitude of the 4-force ${\cal F}$ will depend on the radial distance:
\begin{equation}
PL = PK \ \cos(\alpha) = (PN + NK) \ \cos(\alpha) = (R + v t) \gamma(v)
\end{equation}
and on the length ratio:
\begin{equation}
{\cal F} = {1 \over 2} {Q q \over (PL)^2} \gamma(v) (1 + {v \over c}) = {1 \over 2} {Q q \over \gamma(v) (R + v t)^2} (1 + {v \over c})
\end{equation}
The net parallel component (acting on a segment $s_o$) will be:
\begin{equation}
{\cal F}_{\Vert} = i {1 \over 2} {Q q \over (R + v t)^2} (1 + {v \over c}) (1 - {v \over c})
\end{equation}
This force has to be integrated between B ($t_B = - R/c$) and D ($t_D = R/c$).
\begin{equation}
\sum {\cal F}_{\Vert} = \sum {{\cal F}_{\Vert} \over s_o} s_o = \int {{\cal F}_{\Vert} \over s_o} ds = {i c \over s_o} \int_{t_B}^{t_D} {\cal F}_{\Vert} dt = i {Q q \over \tau_o R c}
\end{equation}
Remarkably, this expression has no dependence on the velocity $v$ of the source. A similar expression, but of a different sign, is obtained for the interaction with the retarded source. The overall change of the rest mass (of the tension in the string) is zero! Thus only on a time scale of $R/c$, where $R$ is the distance between accelerated interacting charges, could someone hope to experimentally detect the variation of the rest mass. 

An essential part of our treatment was the separation of charges into test and source particles. Since even the problem of two particles interacting one with the other through retarded potentials is not tractable, it is hard to believe that our conjecture will ever become a theorem.

\section{The variation of the rest mass is compatible with the equivalence principle}
Consider an infinitesimal SpaceTime volume element, surrounding an infinitesimal length element of a particle's world-line. What happens if we bring a massive body, generating gravitation, in the proximity of our initial particle? The Riemann metric tensor of the space will change, modifying the volume, phenomenologically producing a change in the mass density of the volume element. In the above description we have implicitly considered that the rest mass of the test particle remains constant. But it has to be realized that we have no control over the mass of the particle! This mass {\it cancels} from the equations of motion, as a result of the equivalence principle \cite{einstein}: "[...] the law of the equality of the inert and the gravitational mass is equivalent to the assertion that the acceleration imparted to a body by a gravitational field is independent of the nature of the body." We can alternatively imagine the result of the process of bringing the massive body closer to our test particle as being a change in the rest mass of the particle, while preserving the Minkowski metric, such that we end up with the same change of the mass density.

We have also implicitly assumed that the linear charge density is not changed during the variation of the linear rest mass density. This is because the electric charge is not correlated with inertia, in contrast to mass and energy. Our assumption is also in qualitative agreement with experimental results from Particle Physics which indicate that the charge of a particle can take only a quantized value, while the rest mass is not subject to such a restraint.

\section{A theoretical derivation of the homogenous Maxwell's equations using the Riemannian structure of SpaceTime}
We now generalize the relativistic Lorentz 4-force, such that the radiation reaction be included into the field-strength tensor. We exploit the fact that any 4-force parallel to the 4-velocity can be written as the contraction between a symmetric tensor and the 4-velocity.
\begin{equation}
{\cal F}^{\rm (rad)}_{\alpha} = {2 \over 3} {q^2 \over c^3} |{d^3 \bbox{\cal X} \over d\tau^3} \cdot {d \bbox{\cal X} \over d\tau}| {1 \over - c^2} \delta_{\alpha\beta} {\cal V}_{\beta} = {q \over c} {\cal T}^{\rm (rad)}_{\alpha\beta} {\cal V}_{\beta}
\end{equation}
Consider next two spacetime events, A and B, and a charged particle moving from A to B on any possible smooth path $\Gamma$, restricted only to the conditions that the initial and final velocities be given. Since the direction of the final 4-momentum is given {\it ab initio}, and its magnitude is given by the rest mass (which is {\it uniquely} determined by the Riemann metric tensor at point B, following our analogy, and does not depend on the history of the test particle), we can conclude that the variation of the 4-momentum between A and B is the same regardless of the path followed. For two different paths, $\Gamma_1$ and $\Gamma_2$, we can write
\begin{equation}
\bbox{\cal P}_B - \bbox{\cal P}_A = \int_{\Gamma_1} d\bbox{\cal P} = \int_{\Gamma_2} d\bbox{\cal P}
\end{equation}
The expression of the Lorentz 4-force ${\cal F}_{\alpha} = (q/c) T_{\alpha\beta} {\cal V}_{\beta}$ allows us to write the integrals in (40) as the circulation of the field-strength tensor
\begin{equation}
\int_{\Gamma} d{\cal P}_{\alpha} = \int_{\Gamma} {\cal F}_{\alpha} d\tau = \int_{\Gamma} {q \over c} {\cal T}_{\alpha\beta} {d{\cal X}_{\beta} \over d\tau} d\tau = \int_{\Gamma} {q \over c} {\cal T}_{\alpha\beta} d{\cal X}_{\beta}
\end{equation}
By collecting the integrals in (41) on one side of the equation, we obtain that in general
\begin{equation}
\oint_{\Gamma} {q \over c} {\cal T}_{\alpha\beta} d{\cal X}_{\beta} = 0
\Rightarrow \oint_{\Gamma} {\cal T}_{\alpha\beta} d{\cal X}_{\beta} = 0
\end{equation}
We now introduce Stokes Theorem for Minkowski-space tensors \cite{landau}, in a rather unused form \cite{synge}:
\begin{equation}
\oint_{\Gamma} {\cal T}_{\alpha\beta} d{\cal X}_{\beta} = {1 \over 2} \int df_{\beta\gamma} 
({\partial {\cal T}_{\alpha\gamma} \over \partial {\cal X}_{\beta}} - 
{\partial {\cal T}_{\alpha\beta} \over \partial {\cal X}_{\gamma}})
\end{equation}
where $df_{\beta\gamma}$ are projections of a surface element. Due to the arbitrary nature of the paths $\Gamma_1$ and $\Gamma_2$, from equations (42)-(43) it follows that
\begin{equation}
{\partial {\cal T}_{\alpha\gamma} \over \partial {\cal X}_{\beta}} - 
{\partial {\cal T}_{\alpha\beta} \over \partial {\cal X}_{\gamma}} = 0
\end{equation}
This is the most general condition that the field-strength tensor must satisfy. From (44) we can separate the symmetric and the antisymmetric components:
\begin{equation}
{\partial {\cal T}_{\gamma\alpha}^{(s)} \over \partial {\cal X}_{\beta}} - 
{\partial {\cal T}_{\alpha\beta}^{(s)} \over \partial {\cal X}_{\gamma}} = 
{\partial {\cal T}_{\gamma\alpha}^{(a)} \over \partial {\cal X}_{\beta}} + 
{\partial {\cal T}_{\alpha\beta}^{(a)} \over \partial {\cal X}_{\gamma}}
\end{equation}
From (46) we obtain two more equations by cyclic permutations of the indices ($\alpha \to \beta, \beta \to \gamma, \gamma \to \alpha$). By summing up the three equations thus obtained, the symmetric components cancel, and we end up with the homogenous Maxwell's equations:
\begin{equation}
{\partial {\cal T}_{\alpha\beta}^{(a)} \over \partial {\cal X}_{\gamma}} +
{\partial {\cal T}_{\beta\gamma}^{(a)} \over \partial {\cal X}_{\alpha}} +
{\partial {\cal T}_{\gamma\alpha}^{(a)} \over \partial {\cal X}_{\beta}} = 0
\end{equation}
\section{Loose ends}
The theory we have developed stresses the importance of the world-line of a particle, seen as a string in static equilibrium. How can this picture be reconciled with the Quantum Mechanics elimination of the concept of trajectory? An intuitive answer would be: by considering a {\it vibrating} string. This would of course require the introduction of a fifth dimension, which could be considered as 'the hidden parameter'. Time and Space would no longer be treated on a different footing \cite{debroglie} (time as a parameter and position as an operator), in complete agreement with Special Relativity.

What is more, irreversibility could be introduced as an effect of the 'collapsing of the wave-function' resulted from performing a measurement. The human observer, the human conscience, would thus be inherently recognized as extending into the fifth dimension. The speed $ic$, at which the human conscience is 'traveling' in time, is to be correlated with the velocity of a transverse wave in a string of linear mass density $\mu_o = m_o/s_o$ and under a tension ${\cal T} = ic{\cal P}/s_o = i^2 c^2 m_o/s_o$:
\begin{equation}
v_{transverse \ wave} = \sqrt{{\cal T} \over \mu_o} = ic
\end{equation} 

\section{Conclusions}
We have presented a series of intuitive arguments which suggest that the world-line of a particle be considered as a string in static equilibrium. We have proposed a model of time symmetric action-at-a-distance electromagnetic interaction, which reduces to the classical theory in the case of uniformly moving particles. We have argued that the radiation reaction 4-force should be parallel to the 4-velocity, and that the electromagnetic field-strength tensor should have a symmetric part, which cancels upon averaging. The equivalence principle has allowed us to draw a parallel between our model and General Relativity, and we have derived the homogenous Maxwell's equations as a consequence of the Riemannian structure of SpaceTime.

\begin{figure}
\caption{The equations of motion for a material point particle (a) can be rewritten as the static equilibrium condition for a 4D string (b).}
\label{4Dstring}
\end{figure}

\begin{figure}
\caption{While in the instantaneous action model (a) the force is directed along the line connecting the interacting particles, this is no longer the case when we introduce the retardation time (b). However, a time symmetric interaction model (c) could restore the initial intuitive situation.}
\label{forces}
\end{figure}

\begin{figure}
\caption{Length ratios for a source particle (world-line FI) at rest and a test particle (world-line AB) moving with radial velocity.}
\label{radial}
\end{figure}

\begin{figure}
\caption{Length ratios for a source particle (world-line GJ) at rest and a test particle (world-line OA) moving with arbitrary velocity.}
\label{notradial}
\end{figure}

\begin{figure}
\caption{Calculated in the proper reference frame of the source particle, the real and imaginary components of the advanced and retarded 4-forces add up to the relativistic Lorentz 4-force.}
\label{timesym}
\end{figure}

\begin{figure}
\caption{The variation of the rest mass takes place only on a time scale of R/c, where R is the distance between accelerated interacting charges.}
\label{velchange}
\end{figure}

\begin{figure}
\caption{The test particle is held at rest, at the origin, and the source particle has a radial uniform motion, interrupted only at point A. The total variation of the rest mass of the test  particle, along the segment BD, is zero.}
\label{advanced}
\end{figure}


\begin{references}
\bibitem{davies} Davies P C W, {\it The Physics of Time Asymmetry} (University of California Press, 1977) p.112
\bibitem{ritz} Ritz W, Recherches critiques sur l'\'electrodynamique g\'en\'erale, Ann.Chim.Phys. {\bf 13} (1908) p.145 [Collected Works, p.317]
\bibitem{dirac} Dirac P A M, Classical theory of radiating electrons, Proc.Roy.Soc.London {\bf A167} (1938) p.148
\bibitem{tetrode} Tetrode H, \"Uber den Wirkungszusammenhang der Welt. Eine Erweiterung der klassischen Dynamik., Zeits.f.Physik {\bf 10} (1922) p.317
\bibitem{wheeler1} Wheeler J A and Feynman R P, Classical Electrodynamics in Terms of Direct Interparticle Action, Rev.Mod.Phys. {\bf 21} (1949) p.425
\bibitem{wheeler2} Wheeler J A and Feynman R P, Interaction with the Absorber as the Mechanism of Radiation, Rev.Mod.Phys. {\bf 17} (1945) p.157
\bibitem{landau} Landau L and Lifchitz E, {\it Th\' eorie du Champ} (Moscow: Mir, 1966) p.68, 265, 30 
\bibitem{weyl} Weyl H, {\it Philosophy of Mathematics and Natural Science} (Princeton University Press, 1949) p.116
\bibitem{galeriu} Galeriu C, Electromagnetic interaction between two uniformly moving charged particles: a geometrical derivation using Minkowski diagrams, http://xxx.lanl.gov/abs/physics/0003008
\bibitem{jackson} Jackson J D, {\it Classical Electrodynamics} (New York: Wiley, 1975) p.580, 783
\bibitem{minkowski} Minkowski H, {\it Space and Time}, reprinted in Einstein A, Lorentz H A, Weyl H and Minkowski H, {\it The Principle of Relativity} (Dover, 1952) p.73
\bibitem{rosser} Rosser W G V, {\it Classical Electromagnetism via Relativity} (New York: Plenum Press, 1968) p.118
\bibitem{domina} Moon P, Spencer D E, Mirchandaney A S, Shama U, Mann P J, The Derivation of a New Gaussian Equation for the Force Between Moving Charges from Fundamental Postulates, Phys.Essays {\bf 12} (1999) p.153
\bibitem{pauli} Pauli W, {\it Theory of relativity} (Dover, 1981) p.106, 121, 99
\bibitem{abraham} Abraham M, Zur Elektrodynamik bewegter K\" orper, R.C.Circ.mat.Palermo, {\bf 28} (1909) p.1
\bibitem{einstein} Einstein A, {\it The Meaning of Relativity} (Princeton University Press, 1955) p.57
\bibitem{synge} Synge J L, Schild A, {\it Tensor Calculus} (Dover, 1978) p.275
\bibitem{debroglie} de Broglie L, {\it L'\' electron Magn\' etique} (Paris: Hermann , 1934) p.301
\end{references}
\end{document}